# THE RELATION BETWEEN WEAR AND IRREVERSIBLE ENTROPY GENERATION IN THE DRY SLIDING OF METALS

Hisham A. Abdel-Aal
151 Ottensman hall
Department of General Engineering
University of Wisconsin-Platteville
1 University Plaza, Platteville,
WI, 53818, U S. A
Abdelaah@uwplatt.edu

**ABSTRACT**

This paper examines the relationship between wear and the generation of entropy within the mechanically affected zone, MAZ, of a sliding material. It is postulated that wear is related to irreversible entropy generation within the MAZ. This accumulation takes a place due to the degradation in the ability of the rubbing material to maintain a non-congested flow of the generated entropy. This, in turn, is affected by the nature of the change in the thermal transport properties of the material with respect to the thermal and the mechanical loads acting on the interface. A model, that treats the MAZ as a heat engine in the Carnot sense is presented. The model assumes that the MAZ is a heat engine that transports heat from a high temperature reservoir, represented by the asperity contact layer, to a low temperature reservoir, represented by the sub-contact layer. Consequently an entropy generation source that represents the irreversibilities within the MAZ is defined and a study of the entropy generation is attempted. Wear data, published elsewhere, of two materials, Oxygen Free High conductivity Copper, and Commercially Pure Titanium are analyzed using the developed model. It is found that wear for both materials is correlated to entropy generation, and to the entropy flow within the MAZ. Interestingly, moreover, in view of the contrasting wear trends of the test materials, the relationship of the mass wear rate and the specific wear rate of each material with respect to entropy generation are totally opposite to each other. A common feature between the behavior of the two materials, was found in wear behavior with respect to a so called Ratio of Residual Entropy. This ratio determines wether the system is over or under supplied with entropy. It is found that when the capacity to transport entropy exceeds the entropy supply the mass wear rate increases, and when the entropy transport capacity of the system is exceeded and entropy generation takes place the mass wear rate decreases. This is attributed to the re-establishment of equilibrium within the system.

**Nomenclature**

ACL   Asperity Contact Layer

SRCL  Surface Contact Layer

MAZ   Mechanically Affected Zone

MWR  Mass Wear Rate

NCS   Nominal Contact Surface

NCD   nominal contact datum

RRE   Ratio of Residual Entropy

*ROES*  Rate of Entropy Supply to through the nominal contact surface

SWR   Specific Wear Rate (mass wear rate normalized by thermal load)

*SUCL*  Sub-Contact layer

*WPE*   Wear particle Entropy ( net entropy carried with mass transport)

# 1. Introduction

The phenomena of wear are among the most complex and difficult to model of all mechanical events. This is primarily due to the complex structure of engineering surfaces, severe deformation in the contact region, frictional heat generation, presence of wear debris and contaminants, lubrication, atomic range interactions, chemical reactions on the contact surface etc.,. Due to the challenges that wear offers and its' importance to engineering, it has been a subject of intense scientific and empirical investigations with mechanistic insights that can explain qualitatively a wide variety of wear phenomena quite well [1]. The literature is rich with wear equations where nearly 300 wear-friction models are identified involving an enormous spectrum of materials and operating parameters [2]. In spite of the large number of models found in the literature, no model can predict wear with reasonable accuracy *a priori* based only on material properties and contact information [3].

It is a formidable task, if not impossible, to drive a wear equation from first principles that describes wear satisfactorily and that is truly predictive in nature. Such is the situation because almost all attempts to establish a phenomenological equation to predict wear are centered around mechanical and structural arguments that are concerned with the mechanistic details of sliding events. An alternative approach, that is not common in tribology, is to consider the energetics of wear from a thermodynamic perspective, then attempt to link wear to entropy generation in sliding. Despite its difficulty, this approach upon successful completion allows for the prediction of wear based on intrinsic material properties of the material and not on external influences or sophisticated details of occurrences at the surface (for example). It is recognized that such an approach is not yet possible in its entirety since most of the current wear studies, and the interpretation of wear data, stem from a solid mechanics base.

Wear is in essence an irreversible path dependent process of mass loss. The detachment of wear particles consumes a certain amount of energy. This energy is part of the energy supplied to the material through the work performed on the contacting layers of the rubbing metal by the friction force. So that, wear may be viewed as a sort of a dissipation mechanism to the externally supplied energy. That is, wear particle generation may be considered as an equilibrium restoration mechanism to the contacting layers of the material (the equilibrium of which was disturbed by means of the influence of the frictional force). Complementary to this view, one may further consider that the rate of mass loss (or energy dissipation due to wear particle generation) is an indicator of the irreversibility of the sliding process. Within this context, irreversible thermodynamics seem as a natural candidate to study wear phenomena. In particular, the concept of minimum entropy generation offers a viable tool to study and perhaps predict wear rates. This is because , unlike entropy which is a system property, i.e, doesnot depend on the path of the process, entropy generation, within a thermodynamical system, is a definite positive quantity that depends on the path taken to reach an end state. To this end, one should expect that for each path to mass loss there is an associated entropy generation source. Moreover, in its fundamental form entropy generation is a consequence of the efficiency of heat flow within a system. So that linking entropy generation to mass loss is essentially dependent on identifying the quality of conduction mechanisms within the mechanically affected zone (MAZ) of a rubbing material.

Investigation of wear phenomena based on thermodynamic first principles has received little attention in tribology literature. This is despite that within solid mechanics (which represent the core of wear models formulation) there is a considerable body of literature dealing with phenomenological formulations of material damage based on thermodynamic considerations. Early attempts to investigate friction and wear phenomena within an entropy formulation were undertaken by Klamechi [4-7], who attempted to apply the system-evolution principles developed by Pregogine to tribo-pairs through exergy analysis. Klamechi studied sliding systems based on entropy production for near equilibrium conditions and on conservation of energy for far from equilibrium conditions. One difficulty that Klamechi encountered was to predict conditions necessary for departure from equilibrium for rubbing systems. This difficulty stemmed from his formulation which essentially dealt with entropy change, rather than the rate of entropy generation. Furthermore, in the work of Klamechi of entropy flow due to mass loss was not explicitly expressed so that a direct correlation between wear and entropy was not appare. nt. Nevertheless, Klamechi's work has shown that the steady state thermodynamic response of a tribo-system is determined by the ability of the system to produce entropy. Moreover, it was also shown that when the energy supplied to the system is not compatible with it's energy dissipation capacity, the system response will be unstable. That is, wear is influenced by the ability of the MAZ to maintain efficient, non congested, paths of thermal energy flow. When such a requirement is not satisfied, wear may very well experience a transition in rate and magnitude. There are many indications in tribology literature that this view is meaningful and perhaps promising in describing, qualitatively at least, transition trends of wear regimes. Bian et al [9] recognized that the thermal induced change in a sliding system is an important factor that determines the tribological integrity of a rubbing material. Whence, these authors attempted to link scuffing to frictional heat input. Meng [10], suggested that for every sliding system there is a critical frictional-power input. If operation conditions were such that the frictional power input to the system is less than the critical limit, scuffing (or tribo-failure in general) may be avoided. Whereas, if that input exceeds the critical limit thermal tribo-failure most likely will take place. The ideas of Meng were modified by Jeng [11] who considered that scuffing is the consequence of the disorientation and the desorption of the protective films formed during sliding due to the intensity of frictional heat input exceeding a critical limit.

The conjecture that an equilibrium between frictional heat generation and the heat dissipation was examined by Abdel-Aal [12-16]. It was found that the balance between the heat supplied by friction and the ability to remove heat from the contact zone toward the bulk of the material, away from the contact surface, is dependent on the equilibrium between the conduction ability of the MAZ and thermal diffusion within this layer. It was also found [17-18] that due to the nature of the thermo-mechanical coupling in the contacting layers, the ability of the MAZ to conduct heat away from the surface is likely to be impaired. Such a situation leads to the instantaneous accumulation of t hermal flux causing disequilibrium within the MAZ. The amount of the accumulation correlated to the specific mass loss (rate of mass loss normalized by the friction force) in sliding of Oxygen Free High Conductivity Copper OFHCC [19].

It is constructive at this point to recall that the definition of entropy I sthe heat transfered toa system divided by the absolute temperature at which the transport takes effect, and that entropy generation is given by the net balance between the entropy change (which is related to the heat input to the system) and the entropy transported out of the system. The later is related to the heat transport ability of the MAZ in a wearing metal. Therefore, a correlation between wear and either entropy flow or entropy generation may be hypothesized to exist in analogy to the correlation between wear and thermal flux accumulation.

A correlation between wear (defined as rate of recession of the surface) and entropy flow in lubricated sliding systems was experimentally verified by Ling et al [20,21]. Ling found that the rate of surface recession was proportional to the entropy flow (taken as the sum of heat conducted away from the system at steady state). Furthermore, he used simple arguments to relate the entropy within the system to Archard's wear coefficient. One of the points, not addressed in Ling's work was the effect of total entropy generation on the wear of the system. Since entropy flow represents energy that is already out of the system, then it doesn't contribute to equilibrium within the surface layer. The entropy generated on the other hand, is what remains in the contact layers and therefore it affects thermodynamic equilibrium. The role of this quantity wasn't investigated in the experiments of Ling and coworkers. This paper examines the energetics of wear based on thermodynamic first principles. A relationship between the mass lost through wear and the strength of entropy generation within the MAZ is hypothesized. This hypotheses is verified through the study of wear data for two materials, oxygen free high conductivity copper and pure titanium. The work doesn't aim at balancing the energy spent in each process within the contact layer or to the treatment of chemical reactions at the surface. An inclusion of chemical effects was judged to be out of the scope of the current work. In the absence of exact theory, this work draws on qualitative analysis to gain insight into the physical basis for the hypothesized connection between wear and entropy.

## 2.0   Irreversible Thermodynamics Background

Thermodynamics is based on two fundamental laws: the first law implies conservation of energy and the second law implies that for any process to take place the change in entropy has to be zero or positive, that is entropy can only be created, never destroyed. Irreversible thermodynamics provide a general framework for the macroscopic description of irreversible processes. In irreversible thermodynamics, the so called balance equation for the entropy of a volume element changes with time for two reasons. First, it changes because entropy flows into the volume element. Second, it changes because there is an entropy source due to irreversible phenomena inside the volume element. It is important to note that the entropy source is always a non-negative quantity, since entropy can only be created, never destroyed. For reversible processes the entropy source vanishes.

To relate the entropy source explicitly to the various irreversible processes that occur in a solid system, one needs the macroscopic conservation laws of mass, momentum and energy in local i.e., differential form. The entropy source may then be calculated by using the thermodynamic Gibbs relations, which connects the rate of the change in energy and work. The entropy source has a very simple appearance: it is a sum of terms being a product of a flux characterizing an irreversible process, and a quantity called thermodynamic force, which is related to the non-uniformity of the system [26]. The entropy source strength can thus serve as a basis for the systematic description of the irreversible process occurring in a solid system.

## 2.1 entropy balance

For an irreversible process in which the thermodynamic state of a solid changes from some initial state to a current state, it is assumed that such a process can occur along an imaginary reversible isothermal path, which consists of a two-step sequence [27]. This is the so-called local equilibrium assumption, which postulates that the thermodynamic state of a material at a given point and instant is completely defined by the knowledge of the values of a certain number of variables at that instant. The method of local state implies that the laws that are valid for a macroscopic system remain valid for infinitesimally small parts of it. The process defined in this way will be thermodynamically admissible if, at any instant at evolution, the Clausius-Duhem inequality is satisfied. This condition will be satisfied if the assumption of Fourier type heat conduction is assumed [28].

According to the principles of thermodynamics, two variables temperature, *T*, and entropy , *S*, are used to describe the state of a system. The entropy expresses a variation of energy associated with a variation in the temperature. The entropy of the universe, taken as system plus whatever surroundings are involved in producing the change within the system, can only increase.

The variation of the entropy *dS* may be written as the sum of two and only two, terms:

$$dS = dS_e + dS_i \tag{1}$$

where $dS_e$ is the entropy derived from the transfer of heat from external sources across the boundary of the system, and $dS_i$ is the entropy produced inside the system.

The second law of thermodynamics states that $dS_i$ must be zero for any reversible transformation and positive for irreversible transformation of the system; that is:

$$dS_i \geq 0 \tag{2}$$

The entropy flow term, $dS_e$, on the otherhand doesnot have this restriction and it depends on the interaction of the system with its surroundings. In order to relate $dS_i$ to the various irreversible phenomena occuring inside the system, wer ewrite equations (1) and (2) in local form as:

$$\rho \frac{ds}{dt} = -\operatorname{div} J_s + \gamma \tag{3}$$

and

$$\gamma \geq 0 \tag{4}$$

where:

$$S = \int^{V} \rho s \, dV$$

$$\frac{ds_e}{dt} = -\int^{\Omega} (J_s + \rho s v) \, d\Omega$$

$$\frac{ds_e}{dt} = -\int^{V} v \, dV$$

and *s* is the specific entropy (entropy per unit mass), $J_s$, is the entropy flux and v is the entropy production perunit volume and unit time.

In equilibrium, the total differential of entropy is given by the Gibbs relation that combines the first and the second laws. For an open system this relation takes the form [29],

$$T \, ds = dU + T \sum_k S_k \, d_a n_k + d W_{dis} + \sum_r A_r \, d\xi_r \qquad (5)$$

in which the first two terms of the right hand side relate to the heat and increase in mass due to exchange with the surroundings, the third term to the dissipative effects, and the fourth term to the chemical reactions inside the phase. Differentiating equation (1) with respect to time we may write:

$$\frac{dS}{dt} = \frac{d_e S}{dt} + \frac{d_i S}{dt} \qquad (6)$$

where the *"entropy flow"* is given by:

$$\frac{d_e S}{dt} = \frac{1}{T}\frac{dU}{dt} + \sum_k S_k \frac{d}{dt} n_k \qquad (6\text{-a})$$

and the "entropy production term" is given by:

$$\frac{d_i S}{dt} = \frac{1}{T}\frac{dW_{diss}}{dt} + \frac{1}{T}\sum_r A_r w_r \tag{6-b}$$

It should be noted that $W_{diss}$ ids the available work, namely the work stored in the system during a process. This quantity, unlike the irreversible work, is not a function of the process path. It depends only on the end states of the process for a given environmental temperature. The available work is the maximum amount of work that could be produced by a system between any two given states.

The internal energy rate of change may be expressed as:

$$\rho \frac{du}{dt} = -\nabla \cdot J_q + \sigma:D + \rho r \tag{7}$$

where $r$ represents the internal heat sources (if any) within the solid. Upon substitution of equation (7) in equation (5) and neglecting the chemical reactions term we obtain:

$$\rho \frac{ds}{dt} = -\frac{\nabla \cdot J_q}{T} - \frac{1}{T^2}J_q \cdot \nabla T \frac{1}{T}\sigma:D - \frac{\rho}{T}\frac{d}{dt}W_{diss} + \frac{\rho r}{T} + \sum_k S_k \frac{d}{dt} . n_k \tag{8}$$

From which follows that the expressions for the entropy production are given by:

$$v^o = \frac{1}{T}\sigma:D - \frac{\rho}{T}\frac{d}{dt}W_{diss} - \frac{1}{T^2}J_q \cdot \nabla T + \frac{\rho r}{T} + \sum_k S_k \frac{d}{dt} . n_k \tag{9}$$

Equation (9) represents the entropy production caused by the dissipations. The sum of the first two terms is called the *"intrinsic dissipation"* or mechanical dissipation which consists of plastic dissipation plus the dissipation associated with the evolution of other internal variables. The following two terms are the *"thermal dissipation"* due to the conduction of heat and internal heat source. The last term is the mass change due to interaction with the surroundings.

Since the interest here is solid bodies, then we need only consider the transfer of heat from point to point by conduction then equation (9) would comply with the so called Clausius-Duhem inequality which is another statement of the second law of thermodynamics.

Now, noting that heat conduction within the MAZ entrains a coupled state of thermal and mechanical influences, a modification of the Fourier-Heat equation is necessary. Such modification may be implemented through the Jeffery Heat equation (JHE). This equation may be recast in the Fourier form through the introduction of an apparent thermal conductivity *"K\*"*, given by:

$$K^* = K_o(1 + \beta T) - \frac{3\varepsilon^o K_b K_\varepsilon}{\nabla^2 T} \tag{10}$$

so that the Fourier law of heat conduction may be stated as:

$$J_q = -K^* \nabla T$$

whence the expression of the entropy production source will simplify to:

$$v^o = \frac{1}{T}\sigma:D - \frac{\rho}{T}\frac{d}{dt}W_{diss} - \frac{K^*}{T^2}J_q|\nabla T|^2 + \frac{\rho r}{T} + \sum_k S_k \frac{d}{dt}.n_k \tag{11}$$

Because the entropy change caused by the heat transfer between systems and surroundings has no influence on the degradation of the materials, only the entropy source strength, namely entropy created in the system, should be used as a basis for the systematic description of the irreversible process. Uisng the identity $\sigma{:}\varepsilon - \rho \overset{\circ}{w}_{diss} = \sigma{:}\varepsilon^p$ in equation (11) and assuming no internal heat sources within the system we may write:

$$\begin{aligned} s - s_o &= \int_{t_o}^{t} v^o \, dt \\ &= \int_{t_o}^{t} \frac{\sigma{:}\varepsilon^p}{\rho T} dt + \int_{t_o}^{t} \frac{K^*}{\rho T^2} |\nabla T|^2 \, dt + \int_{t_o}^{t} \sum_k S_k \frac{d}{dt} n_k \end{aligned} \qquad (12)$$

### 3.0 Application to sliding metals

Equation (12) defines the strength of the entropy source as:

$$v^o = \frac{\sigma{:}\varepsilon^p}{\rho T} + \frac{K^*}{\rho T^2} |\nabla T|^2 + \sum_k S_k \frac{d}{dt} n_k \qquad (13)$$

which implies that the strength of the entropy source for an open system is the net balance of three contributions: the work of the tractions, the amount of heat conducted away from the system and the entropy carried in and out of the system due to mass exchange. It is interesting to note that within a sliding system the work of the tractions $\sigma{:}\varepsilon^p$ is equivalent to the work of the friction force. So that, in effect, the work of the plastic tractions can be considered as an amount of heat supplied to the system through it's boundary. This, of course, is not accurately the case, within a strict mathematical sense. Nevertheless, within the frame of this work, and given the nature of the assumptions or approximations implied, the equivalency of the work of the tractions to a boundary supplied heat is fairly admissible.

At this point it is essential to distinguish between two heat quantities, the amount of heat generated at the surface due to friction, $Q_{gen,}$ say and the amount of heat that is conducted through the MAZ away from the surface of contact in the direction of the temperature gradient. The first quantity, enters the MAZ through the contact spots, whereas the second quantity attempts to leave the system and dissipate into the bulk of the material. It is to be noted that the term, dissipation, is used here to indicate the transport of thermal energy from a location of high energy density to ward a location of lower energy concentration (e. g., from the MAZ to the bulk of the material). As such, the usage of the term dissipation of this work engulfs the transport of thermal energy, and the creation of the entropy associated with the flow, to reduce it's intensity and initiate transformation into a different energy form. The third quantity to be identified, is the heat carried out by the wear particles ejected out of the system, $Q_w$. This quantity represents the amount of heat that leaves the system due to generation of wear particles at a given temperature. It is also related to the mass flow within the MAZ. As such, we within the MAZ we encounter a heat supply entering the system through the asperity contacts that is a manifestation of the work done by the friction force and two quantities that are a manifestation of heat flow leaving the system. Naturally, each of these quantities is associated with a ceratin entropy production or transport, and it is the balance between these quantities and their relation to wear particle generation that is subject of this work.

Figure 1(a- through c) presents a schematic illustration of the interaction between the different heat quantities and the entropy associated with them in the MAZ. Figure 1-a, depicts the mechanical forces affecting two bodies in sliding contact, where a normal complying force is acting on the bodies and a tangential friction force, $\mu F$, is acting on the interface. The work of the tangential force results in a quantity of heat that is to be shared by both of the rubbing bodies.

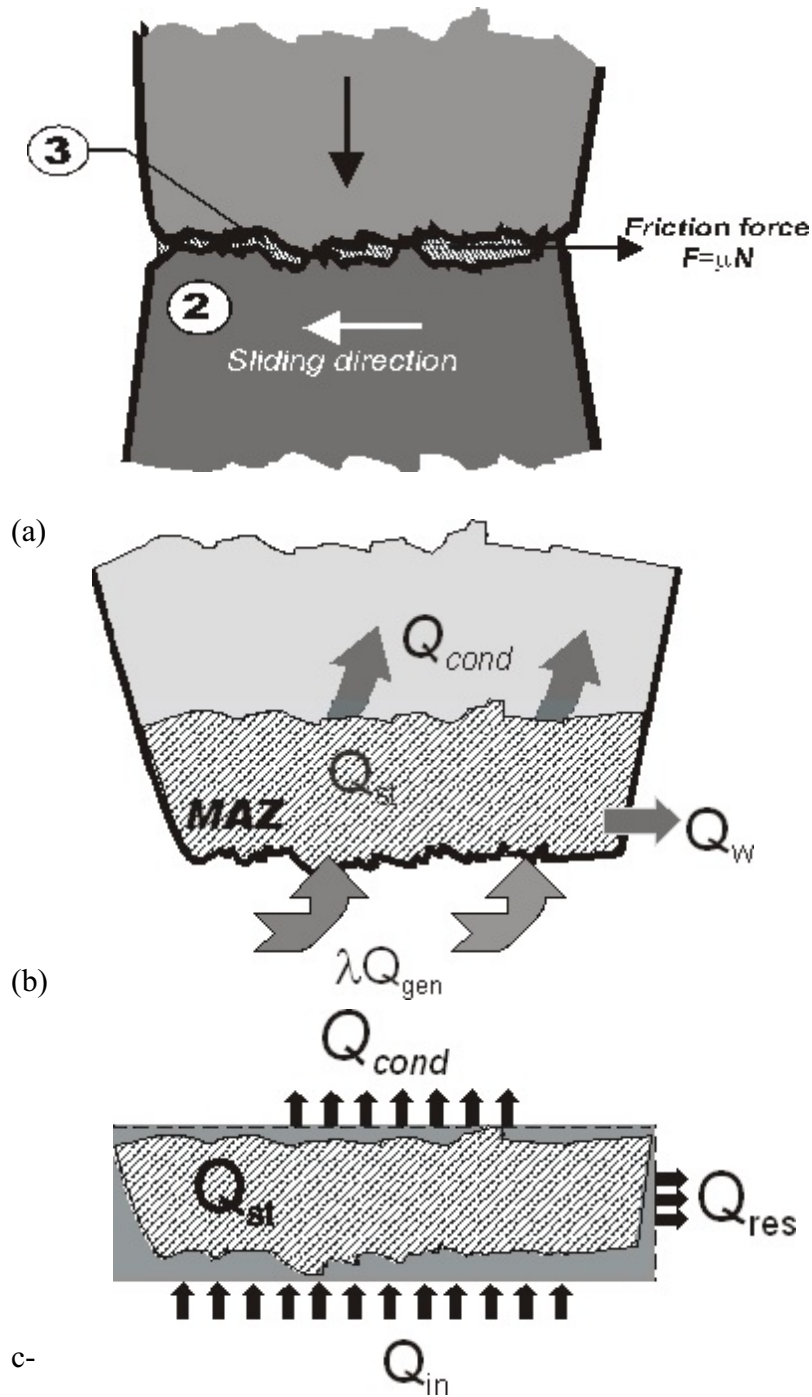

(a)

(b)

c-

Figure 1:  Schematic of thermal interaction within the mechanical affected zone in a sliding material.

A fraction of this thermal load, λ, will be diverted into body 1 (figure 1-b), where it represents the thermal load ($\lambda Q_{gen}$) acting on the contact surface of body-1. Part of this thermal load will be conducted toward the bulk of the material, $Q_{cond}$, another part will be consumed in raising the internal energy within the MAZ, $Q_{st}$, and part will be transferred away by means of the wear particles (mass leaving the system). The system thus reduces to that shown in figure (1-c) where all the thermal quantities are schematically identified.

## 3.1 Strength of the entropy source

The strength for the system shown in figure 1 may be rewritten as:

$$\dot{S}_{gen} = \frac{\lambda \dot{Q}_{gen}}{T} - \frac{\dot{Q}_{cond}}{T} + \sum_{in} \dot{m}_i s_i - \sum_{out} \dot{m}_{out} s_e \qquad (14)$$

The thermodynamic system associated with equation (14) is shown in figure 2 (a through c). It is envisioned that both entropy generation and entropy flow take place due to the exchange of heat and mass between three reservoirs. The first represents the Asperity Contact layer (ACL) within which, the asperities assume the role of ports or inlets to the system. This is where the heat generated due to friction enters the system (figure 2-b). Concurrently, the transport of that heat across the boundary to the surface contact layer (SRCL) is initiated. Thus heat enters the system at the temperature of the ACL, $T_{fl,}$ whereas the initiation of the thermal transport process takes place at the flash temperature $T_{fl}$.

Mass exchange is assumed to take place due to wear particle generation. This is assumed to take place within the asperity contact layer. Consequently, wear particles are ejected from the system at the flash temperature. Due to wear particle generation, a new layer of the SRCL will enter the ACL. The freshly exposed layer will enter the *ACL* at the temperature dominating the *SRCL*.

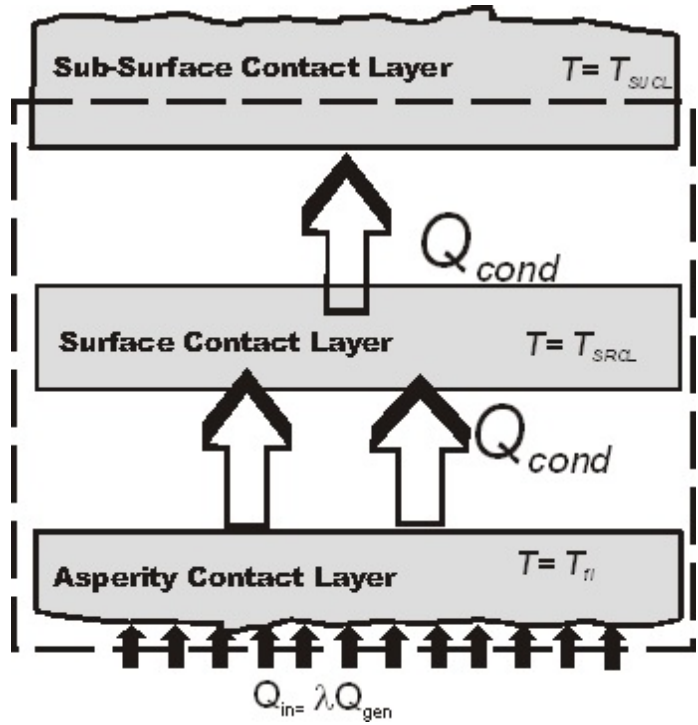

Figure 2-a

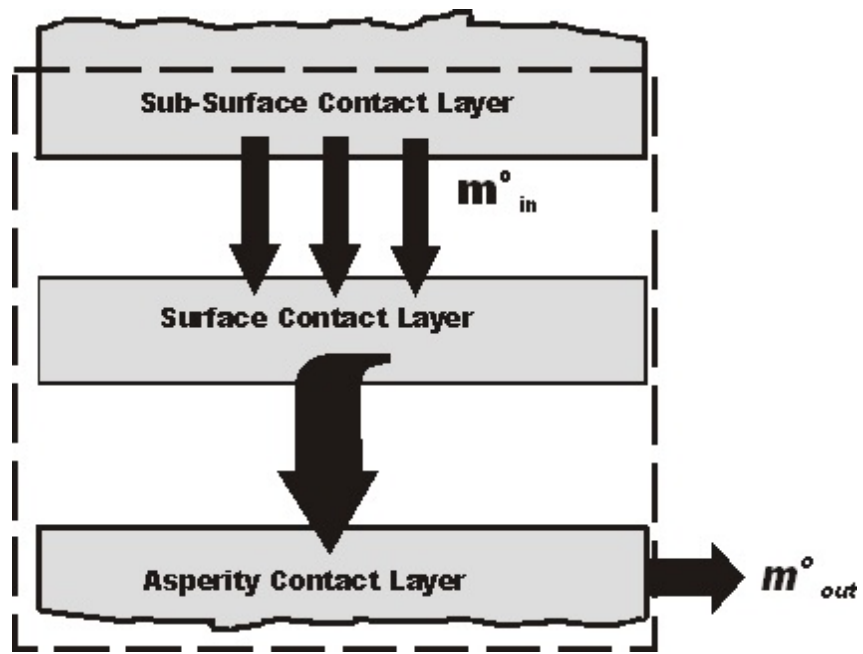

Figure 2-b

figure 2-c

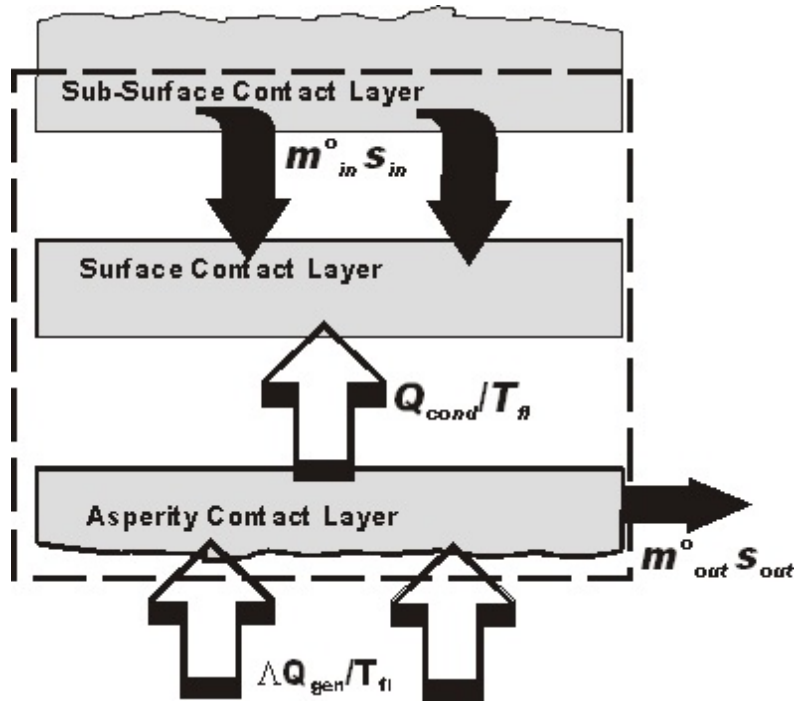

*Figure 2: Thermodynamic model of the Mechanically affected Zone*
*a- Thermal transport, b- mass transport, c- entropy transport*

The recession of the surface with respect to the nominal contact datum, *NCD*, will cause another layer of the material at the opposite end of the system (i.e, at the boundary between the SRCL and Sub-Contact layer, *SUCL*) to enter the system. Thus from this side a fresh layer of material originally at the temperature of the SUCL will enter to the SRCL. As such, the amount of mass entering the system from the SUCL will be equal to the amount of mass ejected out of the ACL. This is equal to the mass of the wear particles. Although the mass entering the system is equal to which is leaving the system, the entropy carried by each quantity is different. This is because the temperatures of the layers supplying the mass within the system are different. This follows from the pattern of heat flow within the system. In order to accommodate heat conduction it is imperative to have,

$$T_{ACL} = T_{Fl} > T_{SRCL} > T_{SUCL} \tag{15}$$

Note that inequality (15), in consistence with the zeroth law of thermodynamics, is what drives the flow of heat within the system and also is what causes the specific entropies of the mass coming into the system to be different than that associated with the mass leaving the system.

Assuming homogenous composition of the mass entering and leaving the system, and assuming a uniform temperature within the different layers of the system, we can write the specific entropies in and out of the system as:

$$\dot{S}_{in} = C \ln\left(\frac{T_{SRCL}}{T_{SUCL}}\right) \tag{16-a}$$

$$\dot{S}_{out} = C \ln\left(\frac{T_{ACL}}{T_{SRCL}}\right) = C \ln\left(\frac{T_{Fl}}{T_{SRCL}}\right) \tag{16-b}$$

Where the values of the of the specific heat, C, are average quantities for the particular limits of the temperature rise in each of the involved layers. That is if $T_{L1}$ and $T_{L2}$ represent the temperature limits of the boundaries of a layer then,

$$C_{T_{L1} \to T_{L2}} = \frac{1}{T_{L1} - T_{L2}} \int_{T_{L1}}^{T_{L2}} C(T) \, dT$$

Equations (14-16) completely characterize the components of contributing to the entropy generation source within the MAZ.

The model represented here is by no means complete. Indeed the mathematical description and the assumptions implied may be enhanced. For example, the model can be applied to a macroscopic entity representing the real area of contact and immediate vicinity. It also can extend to include the effect of oxidation by considering the chemistry of the oxidation process. This will allow for the entropy created by the chemical species. One of the immediate improvements to the quantitative aspects of the model is to consider the composition of the wear debris and characterize their thermal properties especially the heat capacity. It is to be noted however, that the suggested improvements would only enhance the quantitative aspects and not the qualitative conclusions of the model.

**4.0    Results and Discussion**

Model equations (13-16) were applied to published wear data [22-25] for two materials, Oxygen Free High Conductivity Copper (OFHCC) and Commercially Pure Titanium (CPT). Table -1 presents a summary of the conditions under which the wear data were obtained.

**Table-1    Summary of the Experimental Conditions for OFHCC and CPT [22-25]**

| Material | OFHC-Copper | Commercially Pure Titanium |
| --- | --- | --- |
| Test Configuration | Pin-on-Disk | Pin-on-Disk |
| Counter Face | High Purity Alumina | High Purity Alumina |
| Nominal Load (N) | 50 | 50 |
| Sliding Speed (m/s) | 0.1, 0.4, 0.7, 1.5, 2.0, 3.0, 4.0 | 0.1, 0.4, 1.5, 4 |
| Pin Size | 5 mm x 5 mm | 6.0 mm Diameter |
| C. O. F | $0.45 \pm 0.05$ | $0.35 \pm 0.05$ |

Whereas, table -2 represents the thermo-physical properties for the test materials

**Table-2    Summary of the Thermo-physical properties of the materials**

| Material | OFHC- Copper | Pure Titanium |
|---|---|---|
| Nominal Conductivity W/m°K | 401 | 22.1 |
| Specific Heat Capacity (Kj. Kg$^{-1}$°K$^{-1}$) | 345 | 522 |
| Coef. of thermal Expansion 1/°C | 1.65E-05 | 8.9 E-06 |
| Thermal Diffusivity m$^2$/s | 1.16 E-04 | 9.42E-06 |
| Temperature Coef. of conductivity β 1/°C | -1.58E-04 | -4.13E-04 |
| Temperature Coef. of diffusivity δ 1/°C | -0.0003064 | -0.0004697 |
| Density kg/m$^3$ | 8954 | 4500 |

Wear data were originally reported as a rate of recession of the surface (m/m) per unit sliding distance [20-22]. These were converted to a Mass Wear Rate MWR (Kgs$^{-1}$). This, in turn, was also converted to a Specific Wear Rate (SWR) through normalization by the power input to the particular material ($\lambda Q_{gen}$), i.e.,

$$SWR = \frac{\dot{m}_{wear}}{\lambda Q_{gen}} \qquad KgN^{-1}m^{-1}$$

Throughout this section, the behavior of each of MWR and the SWR with respect to each component that contributes to the entropy source (equation 13) is presented. Data pertaining to OFHCC will be denoted by closed circles and those for CPT will be denoted by open circles. Figure 3 (a-d) is a plot of the change in the MWR and the SWR with sliding speed. Figures 3-a and b, pertain to OFHCC, whereas, figures 3-c and d pertain to CPT. It is noted, in figure 3-a, that the MWR generally increases with the increase in the sliding speed. A local minima is noted at a speed of about 1.5 m/s.

a-
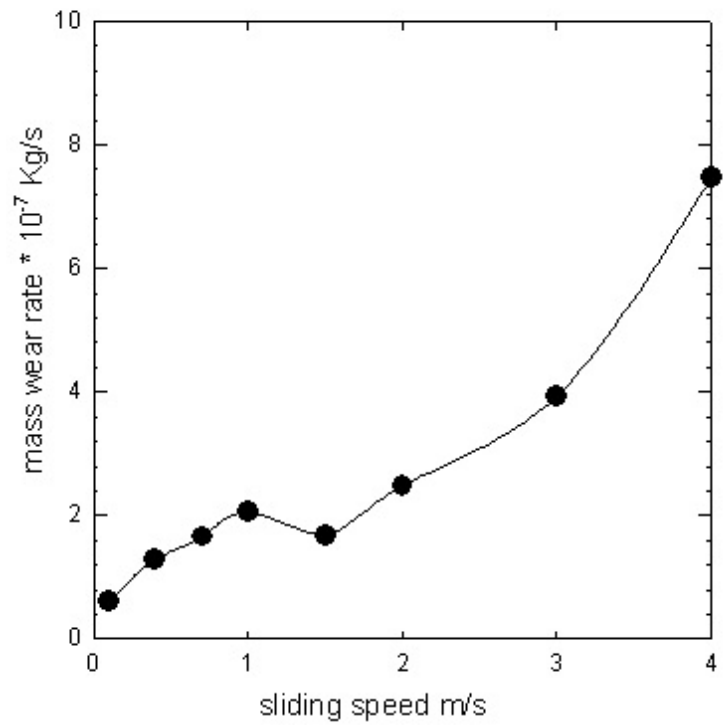

b-
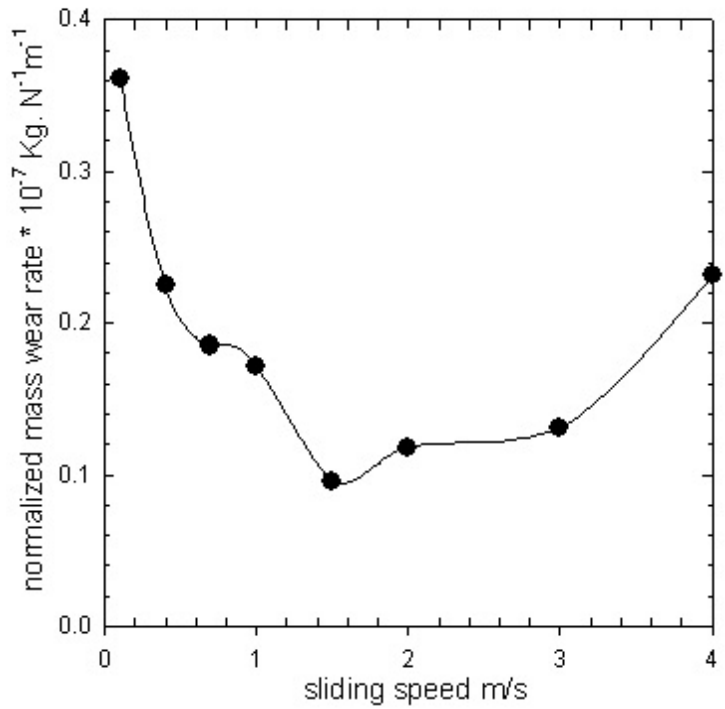

c- 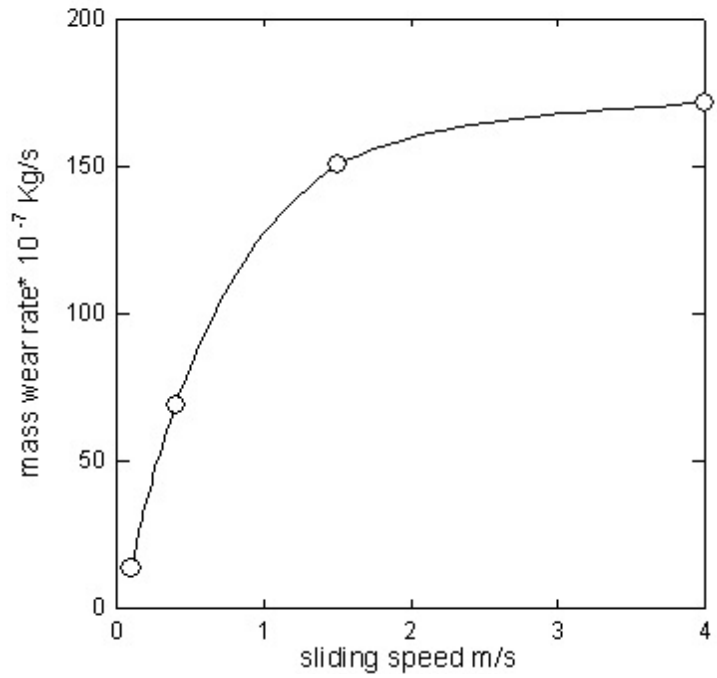

d- 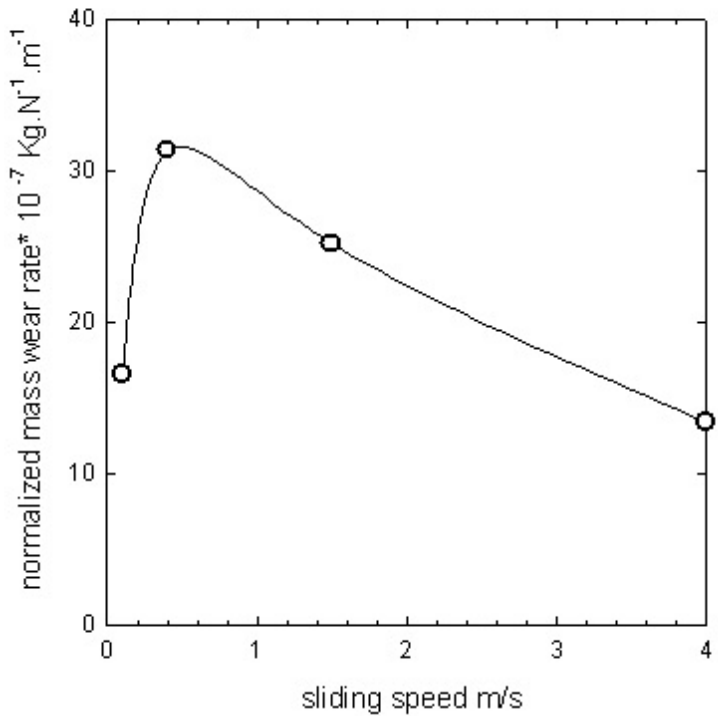

Figure 3:   Variation in the Mass Wear Rate and the Specific Wear rate for OFHCC and CPT with sliding speed

The variation in the SWR can be divided into two parts. The first, represents very slow speeds up to a speed of 1.5 m/s. In this portion of the curve SWR decreases sharply with speed, until it reaches a minimum value at 1.5 m/s. Past this marker the SWR increases with the increase in the sliding speed in a rather uniform manner. However, that increase is not as steep as that exhibited in the low speed portion.

CPT, on the other hand, displays a rather contrasting behavior. Observe that the MWR, figure 3-c, increases sharply with speed until a speed of 1.5 m/s, then past this speed marker, wear almost reaches a plateau. The SWR, in contrast to the behavior of copper, see figure 3-d, increases sharply in the slow peed region until it peaks. Post that peak the SWR decreases mono-tonically with the increase in sliding speed.

The behavior of wear with respect to the rate of entropy supply to the surface ($\lambda Q_{gen}/T_{fl}$) is plotted in figure 4 (a-d). Again, the behavior of both the MWR and the SWR are plotted for each of the test materials. Figure 4-a, is a plot of the MWR for copper against the Rate of Entropy Supply ROES, through the surface for each of the sliding speeds. It is noted that the MWR, in general, increases with the increase in the ROES. Such increase is sharp in the most part. However, it reaches a plateaux at higher speeds. The SWR, displays a contrasting behavior, as it decreases with the increase in entropy supply. This decrease, however, is expected since the amount of thermal energy entering the system at higher speeds is likely to be higher than that entering at lower speeds.

Titanium, figure 4-c, in general displays a similar trend to that of copper. The MWR increases with entropy supply. This is about the only feature that is common between the behavior of the two materials, beyond that the two materials behave in contrast to each other. For example, it is noted that the MWR curve for titanium is concave whereas that of copper is convex. The SWR of titanium

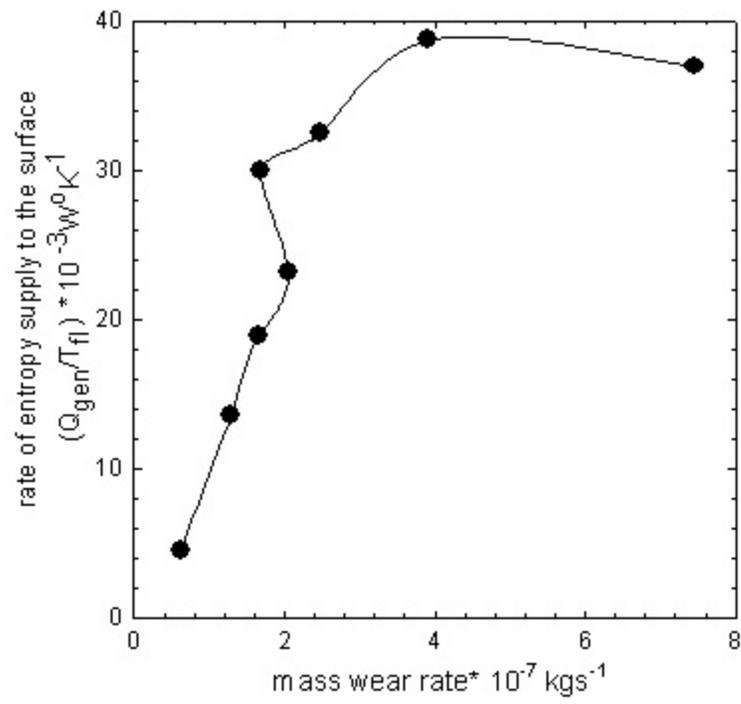

a-

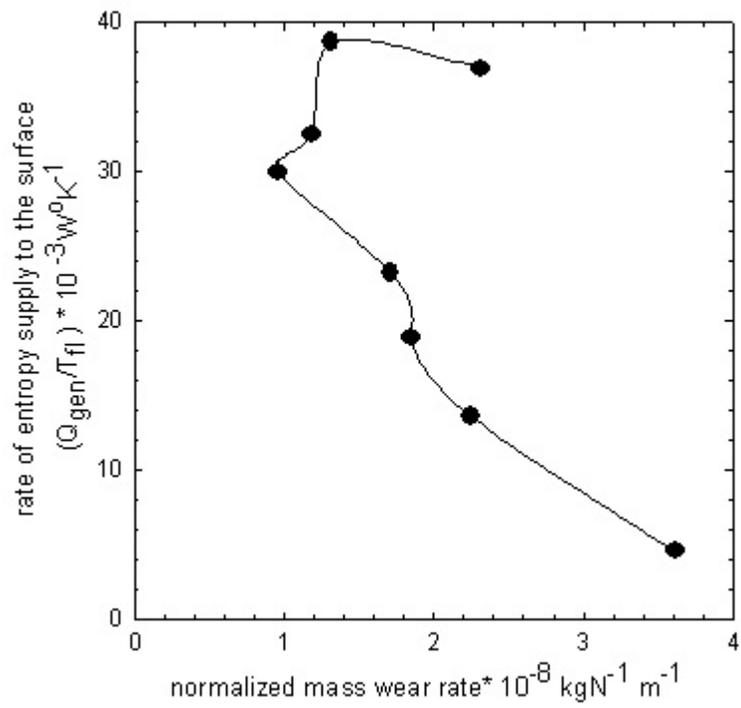

b-

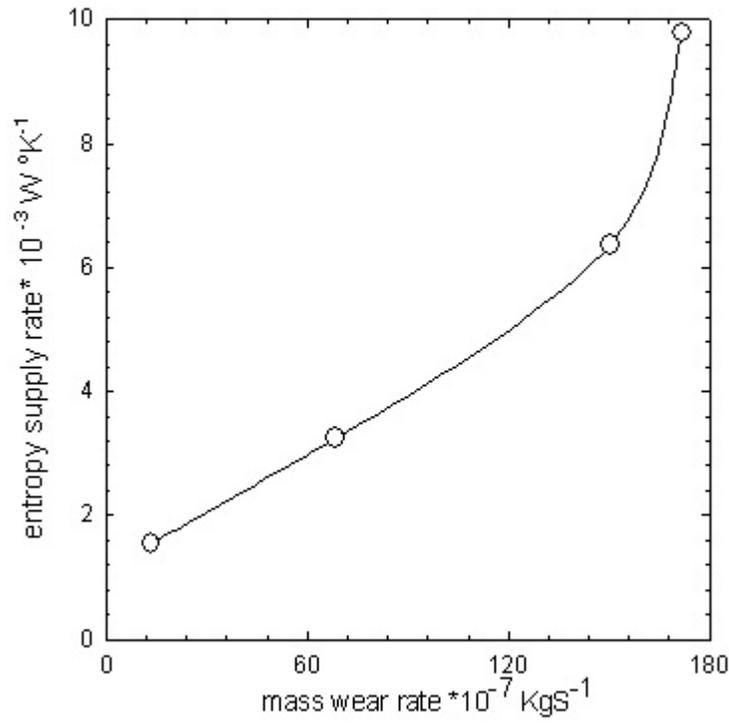

c-

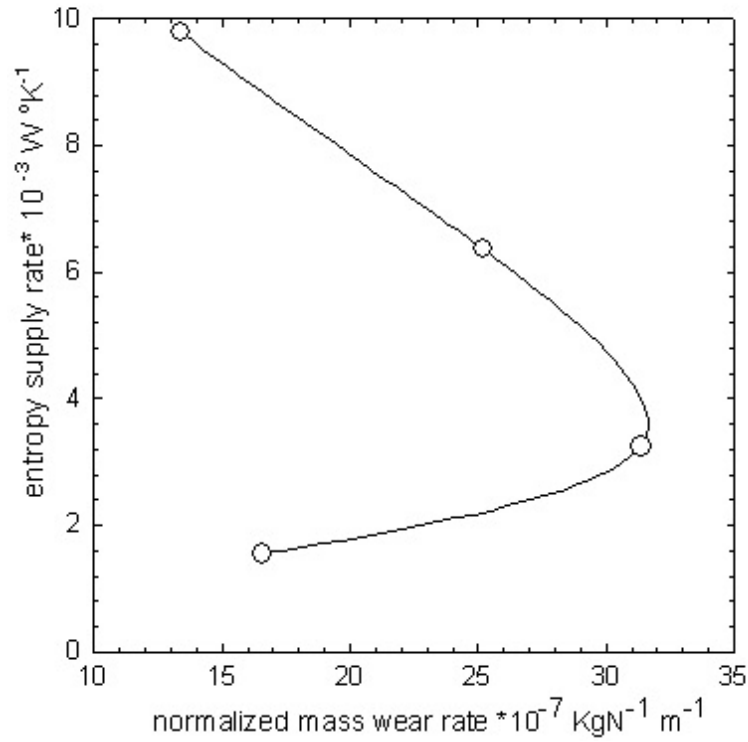

d-

Figure 4    Variation in the Mass wear rate and the Specific wear rate for OFHCC and CPT with the rate of entropy supply through the nominal contact surface

displays an interesting behavior (figure 4-d). In the low speed part an increase in the SWR is noticed with the decrease in the entropy supply up to a critical point (corresponding to 1.5 m/s), the SWR decreases sharply with the decrease in the entropy supply. Again it is noted that the curves for Copper and Titanium display opposite curvature.

The behavior of the SWR and MWR with respect to the rate of entropy flow (a quantity that is directly proportional to the rate of heat removal from the MAZ) are presented in figure 5 (a-d). The negative sign associated with the ROEF indicates that the direction entropy flow is out of the system. It is noted that the MWR for both Copper and Titanium, figure 5 a and c), increase with the increase in entropy flow. This result is in line with the experimental measurement of Ling and co-workers [24,25] for lubricated systems. They also reported increased wear, represented as rate of recession of the surface, with the increase in entropy flow. The SWR however, displays a general tendency to decrease as the entropy flow increases for copper (figure 5-b). When the critical speed of 1.5 m/s is reached the SWR exhibits an increase with a corresponding increase in entropy flow. A contrasting behavior is noted for Titanium (figure 5-d). The SWR increases with speed then reaches a critical point past which, it decreases when the flow of entropy increases. Again opposite curvature trends of the curves are exhibited.

Figure 6 compares the entropy flow due to mass flow out of the system, in the form of wear particles, to the rate of entropy generation, strength of the entropy source calculated from equation 13. Figure (6-a), depicts the variation of the Wear Particle Entropy, *WPE*, with the strength of the entropy source, for copper. It is noted that, in general, the *WPE* increases with the decrease in the strength of the entropy source. Titanium, on the other hand figure 6-b, displays a linear directly proportional relationship with the strength of the entropy source.

a-
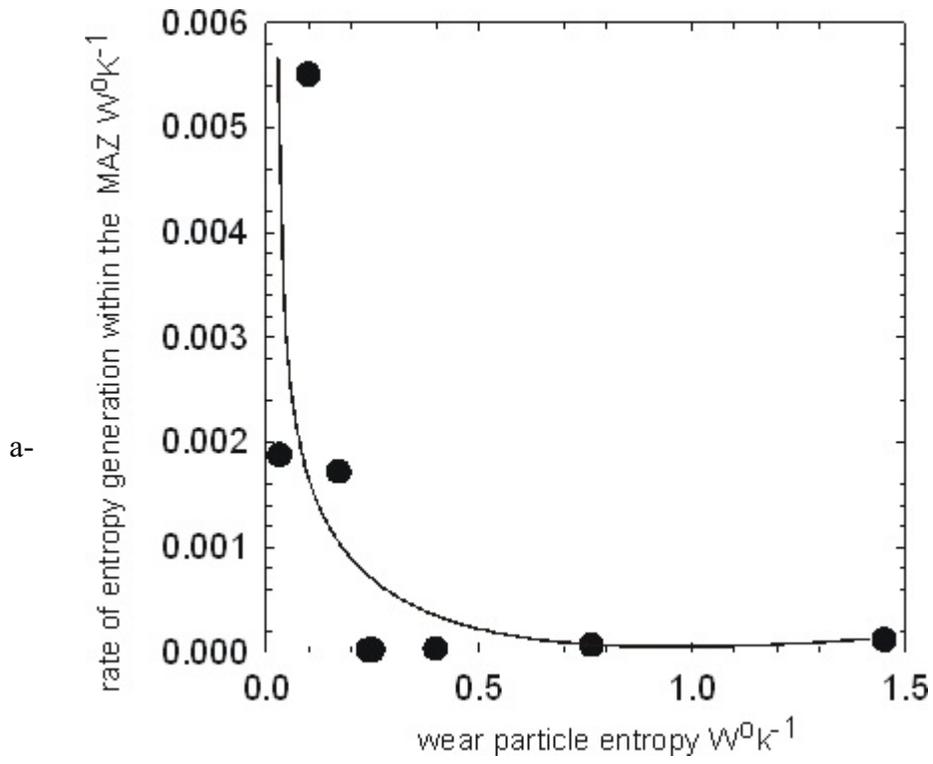

b-
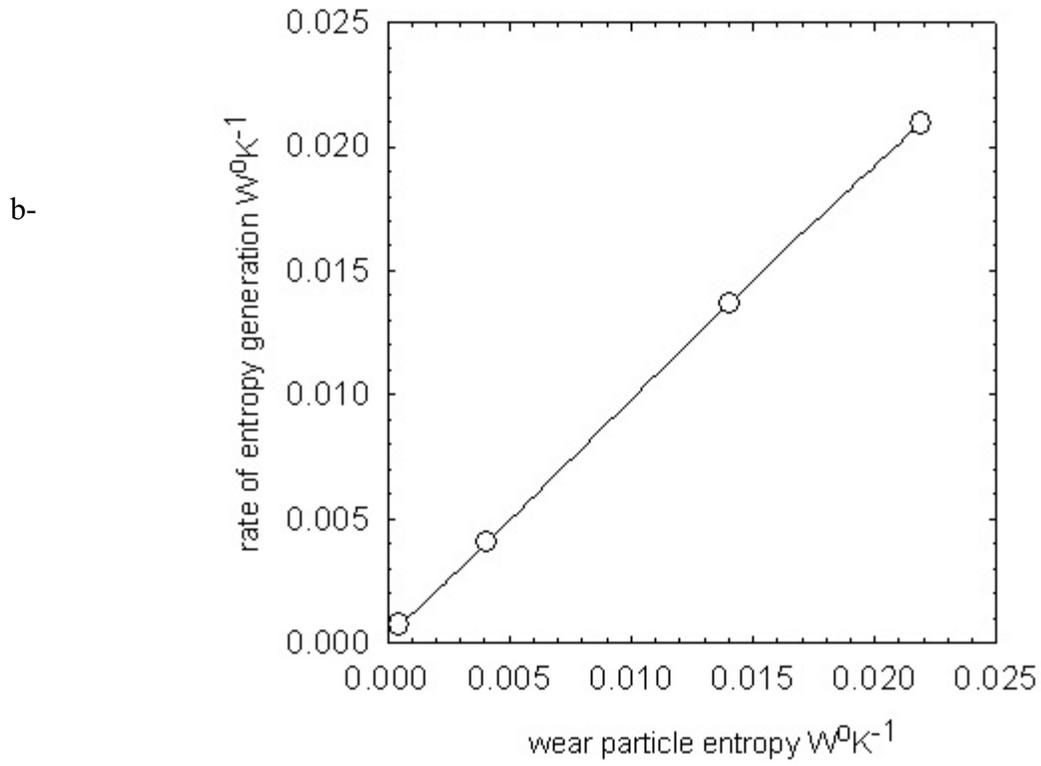

Figure 6: Variation of the entropy carried through mass transport with the strength of the entropy source, a- OFHCC b- CPT

Of interest is to define a ratio of Residual Entropy that defines the ratio of over or under supply of entropy to the MAZ. This can be defined using the first two terms in equation (13). That is by utilizing the net balance between the entropy supply, due to friction work, through the surface and the actual transport of entropy from the ACL to the SCL. So in effect this term is a measure of the effects of the entropy supply due to the external disturbance, the friction force, and the capacity of the MAZ to transport the effects of that disturbance away from the surface. Thus we may write:

$$RRE = \frac{\lambda Q_{gen} - Q_{cond}}{\lambda Q_{gen}}$$

The RRE also may be considered as a measure of irreversibility effects within the surface. This is because the ability of the material to transport heat, and therefore entropy, away from the nominal surface of contact is affected by the effects of the temperature and the state of mechanical strain on the transport properties [23]. The degradation in the transport properties, moreover, is one of the manifestation of irreversibility effects within a tribosystem [21, 29].

Figure 7 (a and b) is a plot of the Ratio of Residual Entropy against the SWR for OFHCC (figure 7-a) and CPT (figure 7-b). For both materials, a transition in the SWR takes place when the RRH changes from a negative sign (indicating under supply of frictional entropy) to a positive sign (indicating over supply of Frictional entropy). This transition is from a lower to a higher SWR for Copper and from a higher SWR to a lower one for Titanium. Both curves, again, display opposite curvature. An interesting behavior, is noted when the MWR for both material is plotted against the RRE as shown in figures (7 c and d). It is noted that the MWR, for both materials exhibits a

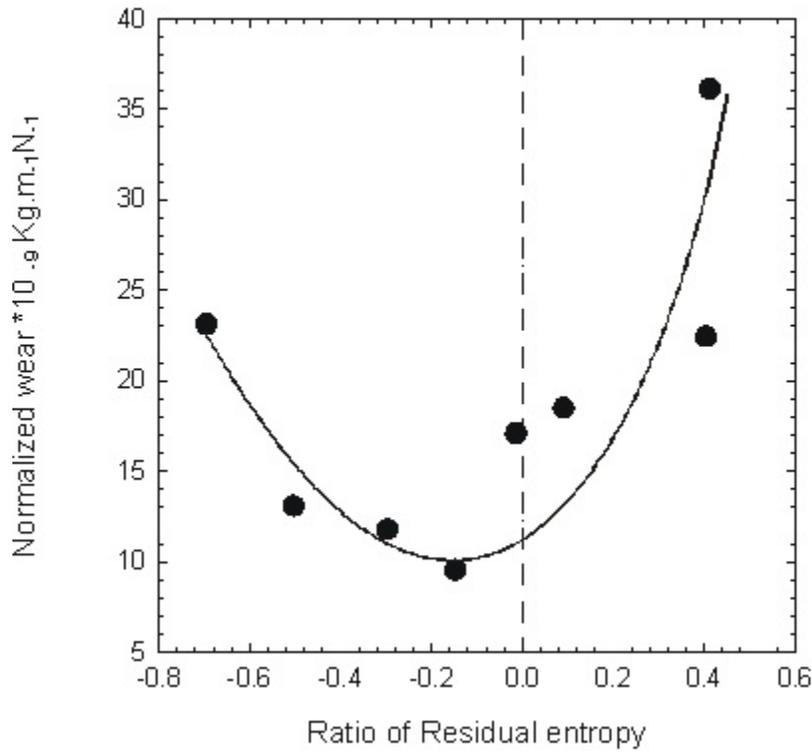

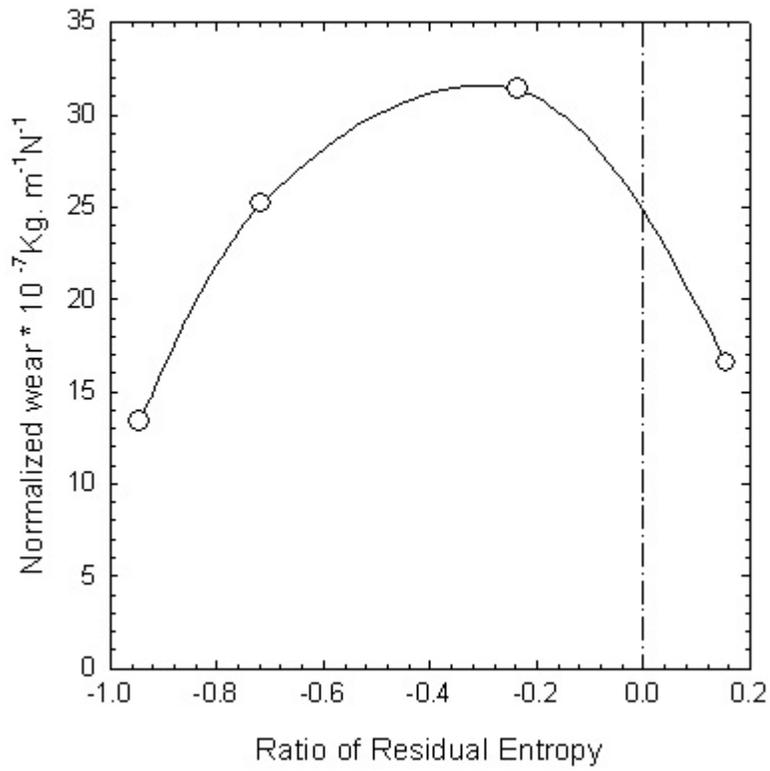

c-
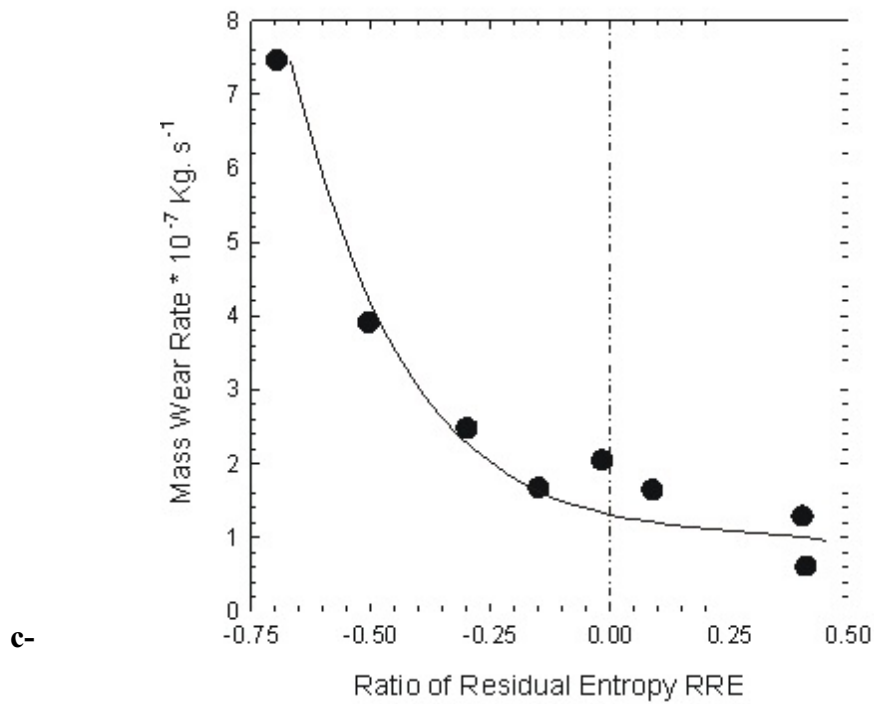

d-
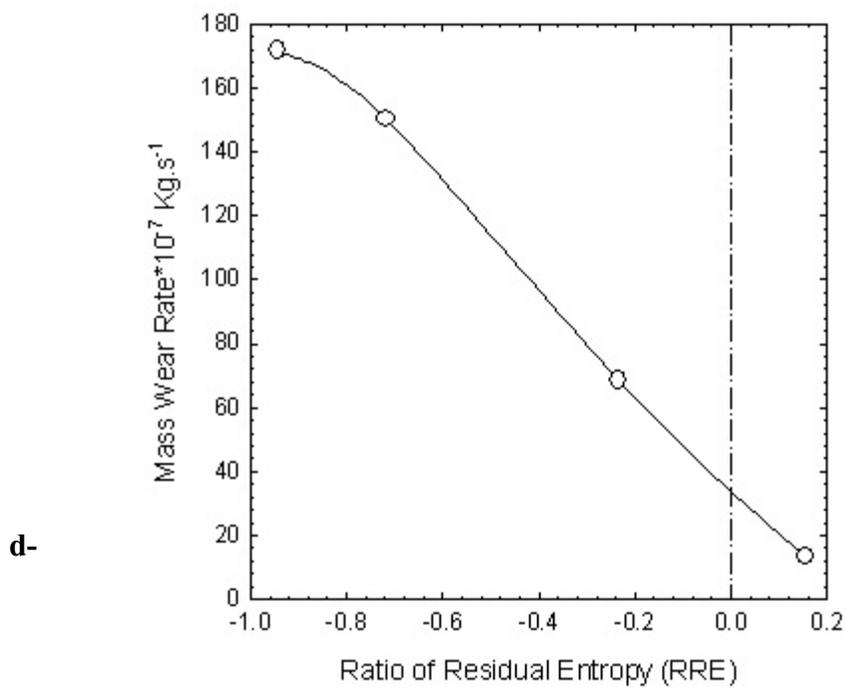

Figure 7: Variation of Wear with the Ratio of Residual Entropy
*a and b, variation in the SWR with the RRE ,*
*c and d: variation in the MWR with RRE*

a transition from a higher wear rate to a lower one upon the change in the sign of the RRE. That is when the entropy supplied to the system doesn't exceed the entropy transport capacity, ETC, of the system the MWR increases. As the amount of entropy supplied approaches the ETC of the system, wear is minimized. This trend continues when there is an excess of entropy within the system. When the entropy supply exceeds the ETC, there will be an entropy accumulation within the system. This will lead the system to evolve into an equilibrium or a quasi-stable (steady state attrition) state. The mechanism of the accumulation and the beneficial bounds of that accumulation may be investigated in detail upon the consideration that the MAZ is a heat engine in the thermodynamic sense of Carnot. Thus the MAZ may be considered a heat engine that transports heat from a high temperature reservoir, represented by the asperity contact layer, to a low temperature reservoir, represented by the sub-contact layer. One can then proceed to calculate the thermodynamic efficiency of such engine and have more insight into the process. However, details of such a process are considered beyond the scope of this work. At any rate, however, the interesting finding remains to be the decrease of mass wear relative to the ETC, and relative to whether or not the ETC of the material is approached or exceeded.